\documentclass[aps,prd,twocolumn,showpacs,preprintnumbers,nofootinbib,amsmath,amssymb,superscriptaddress]{revtex4}

\usepackage{placeins} 
\usepackage{float}
\usepackage{graphicx} 
\usepackage{dcolumn} 
\usepackage{bm} 
\usepackage{amsmath}
\usepackage[normalem]{ulem}
\usepackage[english]{babel}
\usepackage{pdfpages}
\usepackage[a4paper, top=2cm,bottom=1.8cm,left=1.5cm,right=1.5cm]{geometry}
\usepackage{amsmath}
\usepackage[utf8]{inputenc}
\usepackage[T1]{fontenc}
\usepackage{textcomp}
\usepackage{indentfirst}
\usepackage{amssymb}
\usepackage{braket}
\usepackage{epstopdf}
\usepackage{slashed}
\usepackage{epstopdf}
\usepackage[hidelinks]{hyperref}

\newcommand{\beq}{\begin{eqnarray}}
\newcommand{\eeq}{\end{eqnarray}}
\newcommand{\beqnn}{\begin{eqnarray*}}
\newcommand{\eeqnn}{\end{eqnarray*}}

\DeclareMathOperator{\Tr}{Tr}
\DeclareMathOperator{\Index}{Index}
\DeclareMathOperator{\Id}{Id}

\def\spose#1{\hbox to 0pt{#1\hss}}
\def\ltapprox{\mathrel{\spose{\lower 3pt\hbox{$\mathchar"218$}}
 \raise 2.0pt\hbox{$\mathchar"13C$}}}

\begin{document}
\title{Topology via Spectral Projectors with Staggered Fermions}
\author{Claudio Bonanno}
\email{claudio.bonanno@pi.infn.it}
\affiliation{Universit\`a di Pisa, Largo B.~Pontecorvo 3, I-56127 Pisa, Italy}
\affiliation{INFN Sezione di Pisa, Largo B.~Pontecorvo 3, I-56127 Pisa, Italy}
\author{Giuseppe Clemente}
\email{giuseppe.clemente@pi.infn.it}
\affiliation{Universit\`a di Pisa, Largo B.~Pontecorvo 3, I-56127 Pisa, Italy}
\affiliation{INFN Sezione di Pisa, Largo B.~Pontecorvo 3, I-56127 Pisa, Italy}
\author{Massimo D'Elia}
\email{massimo.delia@unipi.it}
\affiliation{Universit\`a di Pisa, Largo B.~Pontecorvo 3, I-56127 Pisa, Italy}
\affiliation{INFN Sezione di Pisa, Largo B.~Pontecorvo 3, I-56127 Pisa, Italy}
\author{Francesco Sanfilippo}
\email{sanfilippo@roma3.infn.it}
\affiliation{INFN Sezione di Roma 3, Via della Vasca Navale 84, I-00146 Roma, Italy}
\date{\today} 

\begin{abstract}
The spectral projectors method is a way to obtain a theoretically well posed definition of the topological susceptibility on the lattice.
Up to now this method has been defined and applied only to Wilson fermions.
The goal of this work is to extend the method to staggered fermions, giving a definition for the staggered topological susceptibility and testing it in the pure $SU(3)$ gauge theory. Besides, we also generalize the method to higher-order cumulants of the topological charge distribution.
\end{abstract}


\maketitle

\section{Introduction}
\label{intro}
The properties related to the existence of field configurations
with non-trivial topology, and the associated non-trivial dependence
on the topological parameter $\theta$, represent some of the most
significant non-perturbative aspects of QCD and QCD-like theories.
Monte Carlo simulations on a lattice are the most natural first
principle tool to investigate such properties, however the fact that
on a discrete space-time homotopy classes are not well defined 
makes the issue non-trivial.
In principle, many definitions of topological charge can be assigned 
in the discretized theory, all consistent with each other
in the continuum limit, however discretization errors 
can be different depending on the choice.

The topological charge
in continuum Yang-Mills theories is defined in terms of gluon fields
as follows:
\beq\label{continuum_charge}
Q = \int d^4x \text{ } q(x) = \int d^4 x \text{ } \frac{1}{64 \pi^2} 
\epsilon_{\mu\nu\rho\sigma}F^a_{\mu\nu}(x)F^a_{\rho\sigma}(x),
\eeq
where $q(x)$ is the topological charge density, and is integer
valued when proper boundary conditions are taken (e.g., periodic 
boundary conditions for a finite box, or vanishing action density 
at infinity).
The index theorem~\cite{Atiyah:1971rm} then relates $Q$ to fermion field 
properties, in particular
\beq
Q = \Index{\slashed{D}} = n_+-n_-,
\label{indexth}
\eeq
where $n_{\pm}$ are, respectively, the number of left-handed and
right-handed zero-modes of the Dirac operator $\slashed{D}$.

The possible lattice discretizations can be divided essentially into 
two different classes: gluonic and fermionic.

Gluonic definitions are based on a straightforward rewriting of
Eq.~(\ref{continuum_charge}) in terms of lattice gauge links.  Despite
having the correct na\"ive continuum limit, they are non-integer
valued and subject to renormalizations induced by ultraviolet (UV)
fluctuations. In particular, correlation functions of $Q_L$ must be
renormalized both additively~\cite{DiVecchia:1981aev} and
multiplicatively~\cite{Campostrini:1988cy} in order to match the
corresponding continuum quantities apart from finite $O(a)$
corrections (where $a$ is the lattice spacing): this is the case for
the topological susceptibility, $\chi \equiv \langle Q^2 \rangle/V$,
as well as for the higher order cumulants of $Q$~\cite{DElia:2003zne}
which enter the coefficients of the Taylor expansion of the free
energy density in $\theta$.  Alternatively, one can make use of
smoothing methods which dampen UV fluctuations of gauge fields while
leaving the global topological background unchanged, thus leading to
an approximately integer valued topological charge: various similar
methods have been proposed, such as cooling~\cite{cooling} or the
gradient flow~\cite{Luscher_wf0, Luscher_wf1}, all leading to
equivalent results~\cite{Bonati:2014tqa, Alexandrou:2015yba}.

Fermionic definitions, being based on a counting of zero modes, are in 
principle better founded, and would be realized in practice
by a simple evaluation of the trace of the 
$\gamma_5$ operator on a basis of eigenvectors
of the Dirac operator. However, also in this case one has 
to face problems related to the difficulty in implementing
fermions with the correct chiral properties on a lattice.
The best approximation is provided by discretizations of the Dirac
operator satisfying the Ginsparg-Wilson relation~\cite{Ginsparg:1981bj}, 
an example 
being the overlap operator~\cite{Neuberger:1997bg}, which 
satisfies an approximate chiral symmetry~\cite{Luscher:1998pqa}, 
leads to a counting of exact zero modes and has been widely used as a tool
to extract the topological content of gauge 
configurations~\cite{DelDebbio:2003rn, DelDebbio:2004ns}.

Alternatively, one can use a more standard fermion discretization,
either Wilson or staggered based. However, in this case zero modes are 
non-exact, or do not have a well defined chirality, or both. 
The counting is not well defined and, if one tries to define $Q_L$
as the trace of the discretized $\gamma_5$ operator, one needs again to 
take into account proper 
renormalizations~\cite{Bochicchio:1985xa, Vladikas_review, smit_and_vink, vink}.
This, however, is not an obstruction, and the method based 
on spectral projectors relies exactly on this strategy.
Indeed, in Ref.~\cite{luscher_1,luscher_2} spectral projectors 
have been used to 
obtain a theoretically well-posed definition of the continuum topological 
susceptibility, which is also easily adaptable for numerical simulations 
on the lattice. In particular, the method has been derived for 
Wilson fermions and successfully tested both in pure 
Yang-Mills~\cite{luscher_3,Cichy_wilson_spectral_quenched} 
and in full QCD~\cite{athenodorou_wilson_spectral_fullQCD}. 
\\

All the methods exposed above, either fermionic or gluonic,
 are theoretically well founded
or designed so as to match the correct definition of homotopy classes
when these become well defined:
as a matter of fact, all methods provide consistent results 
when the continuum limit is taken.
The question about which of the methods
one should adopt can then be answered based on two 
considerations: 
numerical convenience, i.e.~the computational
effort required by a given definition of $Q_L$, 
and the magnitude of residual corrections to the continuum
limit. 

The second issue can be particularly relevant 
for numerical simulations involving light dynamical quarks.
Actually, in this case most of the lattice artifacts 
stem from the discretization of the fermion determinant:
the presence of zero modes should suppress configurations 
with non-zero topological charge, eventually leading to the
absence of $\theta$-dependence in the limit of massless quarks;
however, a lattice discretization with non-exact chiral properties
typically leads to a less efficient suppression, because
of large would-be zero modes, thus leading to somewhat
larger values of the topological susceptibility at finite lattice
spacing. This problem can make the approach to the continuum limit
particularly difficult, both at zero and finite 
temperature~\cite{Bonati:2015vqz,Petreczky:2016vrs,Borsanyi:2016ksw,Burger:2018fvb,Bonati:2018blm}, making
it necessary to perform simulations at lattice spacings much smaller than
those usually adopted in quenched simulations. A possible 
heuristic solution adopted in the recent literature has been 
to reweight gauge configurations by hand, according to the lowest
eigenvalues of the dynamical fermion operator~\cite{Borsanyi:2016ksw}

Even if the above problem is related to the discretized 
path integral measure, rather than to the discretized observable,
it is not inconceivable that the choice of a proper fermion
discretization for the topological charge could ameliorate
the convergence to the continuum, especially if the discretization
matches the one adopted in the measure. This possibility is actually 
supported by a recent study~\cite{athenodorou_wilson_spectral_fullQCD}, 
investigating full QCD 
with twisted mass Wilson fermions, 
in which strongly reduced lattice artifacts
are observed for the zero-temperature topological susceptibility
if a definition based on twisted mass spectral projectors is 
adopted, instead of other standard gluonic definitions.
\\

We can now come to the main point of our study:
we would like to extend the definition of topological 
quantities based on spectral projectors to the case
of staggered fermions, the main motivation being to 
adopt it in ongoing lattice investigations of $\theta$-dependence
in full QCD with staggered fermions~\cite{Bonati:2018blm}.
Most of our discussion will focus on how to properly
define and renormalize the definition of the topological 
susceptibility based on spectral projectors in the case
of staggered fermions; we will also present some numerical
results which however, given the goals of this paper, will
be limited to measurements taken on quenched ensembles;
the case of full QCD ensembles 
will be treated separately in an upcoming work.
In addition, we will also show how 
the spectral projectors method 
can be exploited to define and evaluate cumulants of the 
topological charge higher than just the topological susceptibility.

The paper is organized as follows. In Section~\ref{setup}, after a brief review of the 
method for Wilson fermions, we extend the spectral projectors method 
to the case of 
staggered fermions, deriving spectral expressions for the topological susceptibility and 
for all higher-order cumulants; moreover, in the same section, we also describe the 
numerical strategies we adopted to test spectral projectors on the lattice. 
In Section~\ref{results} we present numerical results for the pure $SU(3)$ gauge theory and 
finally, in Section~\ref{conclusions}, we draw our conclusions and discuss future perspectives. 

\section{$\theta$-dependence via spectral projectors: from Wilson to staggered fermions}
\label{setup}

In this Section, after a brief review of the main ideas 
underlying the method of spectral projectors for Wilson
fermions, we show how they can be extended to the staggered 
case, obtaining a similar expression for the topological susceptibility. We also discuss about the extension
of the method to higher order cumulants and about a practical
way to fix the cut-off scale adopted in the method.

\subsection{Topological susceptibility via spectral projectors: the Wilson case}

As for other definitions of topological charge based on the index
theorem, the starting point is to write it in terms of the trace of the
$\gamma_5$ operator, $Q_0 = \Tr\{\gamma_5\}$. When the trace is taken
over eigenvectors of the lattice Wilson fermion operator, this
definition is subject to a multiplicative renormalization, because
chiral symmetry is explicitly broken by Wilson fermions.  In
particular, making use of non-singlet chiral Ward identities (see
Ref.~\cite{Bochicchio:1985xa} for more details), one can show that the
renormalized charge can be expressed as~\cite{luscher_1,luscher_2}
\beq
Q_L = \frac{Z_S^{(\text{\textit{ns}})}}{Z_P^{(\text{\textit{ns}})}} Q_0,
\eeq
where $Z_S^{(\text{\textit{ns}})}$ and $Z_P^{(\text{\textit{ns}})}$ are, respectively, the renormalization constants 
of the non-singlet scalar and pseudo-scalar fermionic densities
\beq
S_{0,ij} = \bar{\psi}_i \psi_j, \enskip 
P_{0,ij} = \bar{\psi}_i \gamma_5 \psi_j, \quad (i \ne j)\\
S_{ij} = Z_S^{(\text{\textit{ns}})} S_{0,ij}, \enskip 
P_{ij} = Z_P^{(\text{\textit{ns}})} P_{0,ij}, \quad (i \ne j).
\eeq
The correct renormalization factor of the charge can be easily obtained 
from its bare expression once the renormalization constants of the densities 
are chosen according to the non-singlet Ward identities written for Wilson 
fermions (for further details about this topic, we refer to 
Ref.~\cite{Bochicchio:1985xa,Vladikas_review}). 
Besides, note that the ratio $Z_S^{(\text{\textit{ns}})}/Z_P^{(\text{\textit{ns}})}$ is different from 1 at finite lattice spacing because the Wilson operator $D_W$ explicitly breaks chiral
symmetry~\cite{Vladikas_review}.

The renormalization constants $Z_S^{(\text{\textit{ns}})}$ and $Z_P^{(\text{\textit{ns}})}$ can be obtained
by a variety of non-perturbative means. In the spectral projectors approach,
one writes directly their ratio in terms of the 
so-called "spectral sums":
\beq\label{Z_P_over_Z_S_ratio_spectral_sums}
\left(\frac{Z_P^{(\text{\textit{ns}})}}{Z_S^{(\text{\textit{ns}})}}\right)^2 = \frac{\sigma_{k,l}}{\sigma_{k+l}},
\eeq
where
\begin{gather}
\label{wilson_spectral_sum_1}
\sigma_k \equiv \braket{ \Tr \{ (D^\dagger_W D_W)^{-k} \} }, \\
\label{wilson_spectral_sum_2}
\sigma_{k,l} \equiv \braket{ \Tr \{ \gamma_5 (D^\dagger_W D_W)^{-k} \gamma_5 (D^\dagger_W D_W)^{-l} \} },
\end{gather}
since spectral sums can be expressed in terms of density chains, whose renormalization properties are known~\cite{luscher_2}
\begin{gather}
\label{wilson_density_chains_1}
\sigma_k = - \braket{ \Tr\{ P_0^{2k} \} },\\
\label{wilson_density_chains_2}
\sigma_{k,l} = - \braket{ \Tr\{ S_0 P_0^{2k-1} S_0 P_0^{2l-1} \} },
\end{gather}
Note that spectral sums (\ref{wilson_spectral_sum_1}) and (\ref{wilson_spectral_sum_2}) lead to pseudo-scalar density chains (\ref{wilson_density_chains_1}) and (\ref{wilson_density_chains_2}) because of the $\gamma_5$-hermiticity of the Wilson operator: $\gamma_5 D^\dagger_W \gamma_5 = D_W$.

Eq.~(\ref{Z_P_over_Z_S_ratio_spectral_sums}) holds for high enough 
values of $k$ and $l$, but also if the inverse powers of $D^\dagger_W D_W$ are traded for a generic, fast-decreasing function $f(D_W^\dagger D_W)$ (see Ref.~\cite{luscher_2} for more details). Choosing the Heaviside function $f(x)=\theta(M^2-x)$, one can easily evaluate the traces in the spectral basis and obtain 
\beq
\label{wilson_sp_Z_S_Z_P}
\left( \frac{Z_S^{(\text{\textit{ns}})}}{Z_P^{(\text{\textit{ns}})}} \right)^2 = \frac{\braket{\Tr\{\mathbb{P}_M\}}}{\braket{\Tr\{\gamma_5 
		\mathbb{P}_M \gamma_5 \mathbb{P}_M \}}},
\eeq
where $\mathbb{P}_M$ is the orthogonal projector on eigenspaces of the Wilson operator with eigenvalues $\vert \lambda \vert \le M$
\beq
\mathbb{P}_M = \sum_{\vert \lambda \vert \le M} \mathbb{P}_{\lambda}.
\eeq

Analogous considerations can be applied to the fermionic definition 
of the bare charge $Q_0$, which can be rewritten 
in terms of a spectral expression as well:
\beq
\label{wilson_sp_lattice_charge}
Q_0=\Tr\{\gamma_5 \mathbb{P}_M\} \, ;
\eeq
in the continuum, it would suffice to project just 
on the kernel of the operator, however this is of course not true
at finite lattice spacing and for a fermion operator with non-exact 
zero modes.

Finally, one can write the renormalized definition 
of the lattice topological susceptibility via spectral
projectors as follows:
\beq\label{wilson_sp_topo_susc}
\chi_{\text{\textit{SP}}} &=&
\left(\frac{Z_S^{(\text{\textit{ns}})}}{Z_P^{(\text{\textit{ns}})}}\right)^2 \frac{\braket{Q_0^2}}{V} \nonumber \\
&=&
\frac{\braket{\Tr\{\mathbb{P}_M\}}}{\braket{\Tr\{\gamma_5 \mathbb{P}_M \gamma_5 \mathbb{P}_M \}}} \frac{\braket{\Tr\{\gamma_5 \mathbb{P}_M\}^2}}{V}.
\eeq
This definition presents only $O(a)$ or $O(a^2)$ corrections (depending 
on the explicit discretization) if the cut-off mass $M$ is properly 
tuned (i.e.~its renormalized value is kept fixed) as the continuum
limit is taken~\cite{luscher_1,luscher_2,luscher_3}.

An important point, which is worth stressing, is that, because of the 
fast decreasing behavior in the ultraviolet 
of the functions appearing in the spectral sums
(and in particular of the projector $\mathbb{P}_M$),
the above expressions are free of short distance singularities,
so that no further renormalizations are needed apart from the multiplicative
one. In particular, no additive renormalization appears for the topological susceptibility, contrary
to what happens for the standard gluonic definition because of 
contact terms. In this sense, the spectral projectors method has 
some analogies with filtering methods~\cite{Gattringer:2002gn, Bruckmann:2005hy, Bruckmann:2006wf}, in which 
a projection onto the eigenspace of the lowest eigenvectors
of the Dirac operator is used as a smoothing technique.
\subsection{Topological susceptibility via spectral projectors: the staggered case}\label{subsec_stag_sp_chi}
A bare version of the index theorem can be written for the staggered Dirac 
operator $D_{\text{\textit{st}}}$ by just taking into account that, in the continuum, 
it describes
$2^{d/2}$ degenerate flavors of dynamical fermions, where
$d$ is the space-time dimension, so that
the number of zero modes corresponds to $2^{d/2}$ times the topological charge. 
Therefore, one can start from the bare definition $Q_{0\text{\textit{st}}} = (-2)^{-d/2} \Tr\{\Gamma_5\}$, where
$\Gamma_5$ is the staggered version of $\gamma_5$ (more precisely $\Gamma_5 \to \gamma_5 \otimes \Id $ in the continuum limit, see, e.g.,  
Ref.~\cite{smit_and_vink} for an explicit expression).
Chiral symmetry is partially broken also for staggered fermions,
so that $Q_{0\text{\textit{st}}}$ renormalizes multiplicatively as in the Wilson case,
however the renormalization constants are different, since 
the breaking pattern is not the same. Indeed, the staggered lattice action is invariant
under a remnant of the chiral symmetry,
\beq
\psi(x) \rightarrow \Gamma_{55} \psi(x) = (-1)^{\sum_{i=1}^{d}x_i} \psi(x),
\eeq
where $\Gamma_{55} = \gamma_5 \otimes \gamma_5$.

We refer the reader to Refs.~\cite{smit_and_vink,vink} for a detailed
discussion of the anomalous Ward identities for staggered fermions. The 
final result for the renormalized staggered charge, which has been obtained writing the Witten-Veneziano equation starting from the renormalized singlet axial Ward identity, is the following
\beq
Q_{\text{\textit{st}}} = \frac{Z_P^{(\text{\textit{s}})}}{Z_S^{(\text{\textit{s}})}} Q_{0\text{\textit{st}}}
\label{q_stag_ren}
\eeq
where the constants $Z_S^{(\text{\textit{s}})}$ and $Z_P^{(\text{\textit{s}})}$ appear in
the inverse order with respect to Wilson fermions, 
and refer respectively to the scalar and pseudo-scalar flavor-singlet (compared
to non-singlet in the Wilson case) 
 bare densities, $S_0$ and $P_0$:
\beq
S_0 = \bar{\psi} \psi, \enskip
P_0 = \bar{\psi} \Gamma_5 \psi,
\label{scalar_pseudoscalar}
\eeq
meaning that the corresponding renormalized quantities read
\footnote{Note that our notation differs from the one employed in \cite{smit_and_vink,vink}. The bare singlet scalar and pseudo-scalar densities are written as $S_1$ and $P_1$, cf.~Eqs.~(2.5) and (2.6) in \cite{smit_and_vink}, and their renormalization constants are written in terms of the quark mass one $m_R=Z_m^{-1} m$ as $Z_S^{(\text{\textit{s}})}=Z_m k_S^1$ and $Z_P^{(\text{\textit{s}})}=Z_m k_P^1$, cf.~Eqs.~(4.3) in \cite{smit_and_vink}. The constant $k_S^1$ can be taken equal to 1 while $k_P^1 \equiv k_P$, thus $Z_P^{(\text{\textit{s}})}/Z_S^{(\text{\textit{s}})}=k_P$, which is the renormalization constant of the fermionic topological charge appearing in Eq.~(5.24) of \cite{smit_and_vink} and also in \cite{vink}.}
\beq
S = Z_S^{(\text{\textit{s}})} S_0, \enskip
P = Z_P^{(\text{\textit{s}})} P_0.
\eeq
In the staggered case the ratio $Z_S^{(\text{\textit{s}})}/Z_P^{(\text{\textit{s}})}$ can be computed in terms of the staggered spectral sums
\beq
\sigma_k \equiv \braket{\Tr\{(D_{\text{\textit{st}}}^\dagger D_{\text{\textit{st}}})^{-k}\}} = \braket{\Tr \{ S_0^{2k} \}   },
\eeq
and
\beq
\sigma_{k,l} &\equiv& 
\braket{\Tr\{\Gamma_5 (D_{\text{\textit{st}}}^\dagger D_{\text{\textit{st}}})^{-k} \Gamma_5 (D_{\text{\textit{st}}}^\dagger D_{\text{\textit{st}}})^{-l}\}} \nonumber \\ &=& 
\braket{ \Tr \{P_0 S_0^{2k-1} P_0 S_0^{2l-1} \} },
\eeq
through the ratio
\beq
\left(\frac{Z_P^{(\text{\textit{s}})}}{Z_S^{(\text{\textit{s}})}}\right)^2= \frac{\sigma_{k+l}}{\sigma_{k,l}} \, .
\label{spectral_ratio_stag}
\eeq
We note that also in this case the ratio of spectral sums is inverted
with respect to Eq.~(\ref{Z_P_over_Z_S_ratio_spectral_sums}) for Wilson
fermions. This is due to the fact that we are dealing with singlet, rather
than non-singlet, densities.

Following the same line of reasoning of Ref.~\cite{luscher_2}, 
one can trade also in this case the inverse powers of 
$D^\dagger_{\text{\textit{st}}} D_{\text{\textit{st}}}$ for a fast-decreasing function,
in particular for $\mathbb{P}_M$, 
where now $\mathbb{P}_M$ is the orthogonal projector 
on eigenspaces of the Dirac operator with purely imaginary 
eigenvalues $-i \lambda$ such that $\lambda^2 \leq M^2$.
That leads to the following expression for the ratio
\beq\label{stag_sp_Z_S_Z_P}
\left(\frac{Z_P^{(\text{\textit{s}})}}{Z_S^{(\text{\textit{s}})}}\right)^2 = \frac{\braket{\Tr\{\mathbb{P}_M\}}}{\braket{\Tr{\{\Gamma_5 
\mathbb{P}_M\Gamma_5 \mathbb{P}_M\}}}},
\eeq
and finally, using also in this case a spectral projector definition
for the bare topological charge
\beq\label{stag_sp_lattice_charge}
Q_{0\text{\textit{st}}}=(-2)^{-d/2} \Tr\{\Gamma_5\mathbb{P}_M\} \, ,
\eeq
we obtain the following expression for the topological susceptibility of staggered fermions in $d$ dimensions:
\beq
\label{stag_sp_topo_susc}
\chi_{\text{\textit{SP}}} &=& \left(\frac{Z_P^{(\text{\textit{s}})}}{Z_S^{(\text{\textit{s}})}}\right)^2 
\frac{\langle Q_{0\text{\textit{st}}}^2\rangle}{V} \nonumber \\
&=& \frac{1}{2^d} 
\frac{\braket{\Tr\{\mathbb{P}_M\}}}{\braket{\Tr{\{\Gamma_5 \mathbb{P}_M\Gamma_5 
\mathbb{P}_M\}}}} \frac{\braket{ \Tr{\{\Gamma_5 \mathbb{P}_M\}^2 } }}{V}
\eeq
which coincides with Eq.~(\ref{wilson_sp_topo_susc}) apart from the 
factor $2^{-d}$, related to the taste degeneration, as previously explained.
Also in this case, the same considerations apply regarding the fast decreasing
behavior of the projector in the ultraviolet, leading to the absence of 
short distance singularities and additive renormalizations.

As for Wilson fermions, the cut-off mass $M$ appears as a free parameter
of the definition. If the zero modes were exact, one could take
$M$ arbitrarily small. Exact zero modes are obtained for 
the overlap version of the staggered operator~\cite{Adams:2009eb,Azcoiti:2014pfa},  
however, for the standard staggered operator
$D_{st}$ they are shifted by lattice artifacts because of the 
explicit breaking of the chiral symmetry~\cite{vink}, so that
one must extend the sum up to a certain cut-off eigenvalue $M$, 
keeping its renormalized value fixed as the continuum limit is 
approached (see Subsection~\ref{how_to_set_cut_off} for more details).

\subsection{Higher-order terms of the $\theta$-expansion via spectral 
projectors}
The $\theta$-dependence of the vacuum energy (free energy) density can be parametrized around $\theta=0$ as follows~\cite{topo_report}:
\beq\label{theta_dep_vacuum_energy}
f(\theta)\equiv E(\theta) - E(0) = \frac{1}{2} \chi \theta^2 \left(1+\sum_{n=1}^\infty b_{2n} \theta^{2n} \right)
\eeq
where the $b_{2n}$ coefficients, which parametrize the corrections to the quadratic behavior of $f(\theta)$, are defined as:
\beq\label{cont_expression_b_2n}
b_{2n} \equiv (-1)^n\frac{2}{(2n+2)!} \frac{\braket{Q^{2n+2}}_c}{\braket{Q^2}},
\eeq
where $\braket{Q^k}_c$ denotes the $k^{\text{th}}$-order cumulant of the probability distribution $P(Q)$.

These quantities can be computed in terms of spectral projectors as well,
exploiting in particular the fact that, because of the absence of short
distance singularities, only multiplicative renormalizations have to 
be taken into account. In particular, following the same line of thoughts of
the topological susceptibility, it is easy to prove the
following general expression  
\beq\label{general_expression_stag_sp_b_2n}
b_{2n}^{\text{\textit{SP}}} &=& 
(-1)^n\frac{2}{(2n+2)!} \left(\frac{Z_P^{(\text{\textit{s}})}}{Z_S^{(\text{\textit{s}})}}\right)^{2n} \frac{\braket{Q_0}^{2n+2}_c}{\braket{Q_0}^2} \nonumber \\
&=& \frac{(-1)^n}{2^{dn}} \frac{2}{(2n+2)!} \left(\frac{\braket{\Tr\{\mathbb{P}_M\}}}{\braket{\Tr{\{\Gamma_5 \mathbb{P}_M \Gamma_5 \mathbb{P}_M \}}}}\right)^{n} \cdot \nonumber \\
& & \, \ \ \ \ \ \ \cdot \frac{\braket{ \Tr{\{\Gamma_5 \mathbb{P}_M\}^{2n+2} } }_c }{\braket{ \Tr{\{\Gamma_5 \mathbb{P}_M\}^2 } }}.
\label{b_2n_stag_spectr_proj}
\eeq
The above expression has been written explicitly for the case of
staggered fermions but, of course, the final expression holds for
Wilson fermions as well, after omitting the factor $2^{-dn}$, again
related to taste degeneracy.

\subsection{Numerical implementation and remarks on the choice of the cut-off mass $M$}
\label{how_to_set_cut_off}

In the case of Wilson fermions, different strategies have been adopted 
in the literature for the evaluation
of the traces appearing in Eq.~(\ref{wilson_sp_topo_susc}), either by
means of noisy estimators~\cite{luscher_3,Cichy_wilson_spectral_quenched}, 
or by an explicit computation,
configuration by configuration, of all relevant eigenvectors of the 
Dirac operator entering the traces~\cite{athenodorou_wilson_spectral_fullQCD}.
In our numerical implementation
we have followed the second strategy, i.e.~we evaluated the traces 
appearing in Eqs.~(\ref{stag_sp_topo_susc}) and 
(\ref{b_2n_stag_spectr_proj}) expressed the projector 
$\mathbb{P}_M$ through the eigenvectors of $D_{\text{\textit{st}}}$:
\beq\label{projector_M}
\mathbb{P}_M = \sum_{\vert \lambda \vert \le \lambda_{\text{\textit{max}}}} \sum_{\lambda_i = \lambda} u_i u_i^{\dagger}.
\eeq
limiting 
the sum over eigenvalues up the threshold
$\lambda_{\text{\textit{max}}} = a M$.
The relevant quantities entering the expressions for the topological 
susceptibility and the higher order cumulants are then
\begin{gather}
\label{trace_nu}
\Tr\{\mathbb{P}_M\} = \nu(M) \, ,\\
\label{sp_trace_1}
\Tr\{\Gamma_5 \mathbb{P}_M\} = \sum_{\vert\lambda_i\vert \le \lambda_{\text{\textit{max}}}} u_i^\dagger \Gamma_5 u_i \, , \\
\label{sp_trace_2}
\Tr\{\Gamma_5 \mathbb{P}_M\Gamma_5 \mathbb{P}_M\} = \sum_{\vert\lambda_i\vert, \vert\lambda_j\vert\le \lambda_{\text{\textit{max}}}} \vert u_j^\dagger \Gamma_5 u_i \vert^2 \, ,
\end{gather}
where $\nu(M)$ is the number of eigenvalues with $\vert \lambda \vert 
\leq a M$.
\\

As already explained in Subsection~\ref{subsec_stag_sp_chi}, 
the choice of the cut-off mass 
$M$ is irrelevant in the continuum limit, 
since index theorem states that only zero-modes 
contribute to topology. However, 
corrections to the continuum limit do depend on it, and 
it is possible to show that  
lattice artifacts are $O(a^2)$ if the renormalized 
value of the cut-off mass, $M_R$, is kept fixed
as the lattice spacing is varied~\cite{luscher_2,smit_and_vink}):
\beq
\chi_{\text{\textit{SP}}} = \chi + c(M_R)a^2 + O(a^4),
\eeq
where $\chi$ is the continuum value of the topological susceptibility.

Therefore, one has to tune $M$ as a function of $a$ in order
to keep $M_R$ fixed. Most of the following discussion applies
specifically to the case of staggered fermions, for which the 
mass renormalizes as follows~\cite{smit_and_vink}
\beq
M_R = {Z_S^{(\text{\textit{s}})}}^{-1} M
\eeq
where $Z_S^{(\text{\textit{s}})}$ is the renormalization constant of the singlet scalar density 
mentioned above. This quantity is not accessible separately in terms
of spectral sums, however in many lattice studies it is already known
by other means. For instance, in numerical simulations of full QCD
performed on a line of constant physics, one already tunes the 
bare quark masses as a function of the lattice spacing so as to keep
the physics, hence the renormalized quark masses, unchanged: in these
cases it will suffice to keep the ratio of $M$ to any of the bare quark 
masses unchanged as the continuum limit is approached.

However, in the present study we consider as a numerical test-bed
the case of the pure gauge theory at zero temperature, 
for which the strategy above cannot be applied. 
In this case, trying to avoid a direct computation 
of $Z_S^{(\text{\textit{s}})}$ for each lattice spacing, we have devised the 
following strategy. The number of eigenmodes which are 
found below a given threshold scales proportionally to 
the total lattice volume, i.e.~the density of eigenmodes
with $|\lambda| < M$, $\nu(M)/V$, is a constant as the thermodynamical 
limit $V \to \infty$
is approached. Moreover, if the renormalized 
threshold $M_R$ is kept fixed, the density of eigenmodes
is expected to be independent of the lattice spacing.
This is supported by leading order
chiral perturbation theory and the Banks-Casher relation:
in the large-volume and chiral limit, and for small enough $M$,
one has~\cite{luscher_2}
\beq\label{eq_for_cut_off_M}
\frac{\braket{\nu(M)}}{V} = \frac{2}{\pi} \Sigma M
= \frac{2}{\pi} \Sigma_R M_R \, , 
\quad \Sigma = -\braket{\bar{\psi}\psi},
\eeq
where $\Sigma$ is minus the chiral condensate in the thermodynamic and chiral limit, and we have used the fact that $Z_S^{(\text{\textit{s}})}$ is the renormalization
constant for both the singlet scalar density and the inverse mass,
so that $\Sigma M$ is a renormalization group invariant quantity.

Therefore, our prescription in the following will be to keep
the bare quantity $\braket{\nu(M)}/V$ fixed (with the space-time
volume expressed in physical units) in order to 
maintain $M_R$ constant as the lattice spacing is changed.

\section{Numerical tests in the pure gauge theory}
\label{results}
In order to test the definition of topological quantities via
staggered spectral projectors, we have considered the pure 
$SU(3)$ Yang-Mills theory.
Configurations have been 
generated using the standard Wilson plaquette action for $N_c=3$:
\beq
S_L = - \frac{\beta}{N_c} \sum_{x, \mu>\nu} \Re \Tr \{\Pi_{\mu\nu}(x)\}.
\eeq
and a standard local algorithm consisting of a 4:1 
mixture of over-relaxation and over-heatbath.
For simulations at zero temperature, we have considered
4 different lattice spacings, 
corresponding to $\beta=\{5.9, \text{ } 6.0, \text{ } 6.125, \text{ } 6.25\}$,
and symmetric lattices $L \times L \times L \times L$ with 
$L$ in the range 1.2 - 1.8~fm (see Table~\ref{tab_chi_vs_a}), 
which for the pure gauge theory
is large enough to ensure the absence of significant finite size effects,
in particular for topological quantities. In the 
following we will express physical quantities, as well as the 
lattice spacing, in terms of the Sommer parameter $r_0 \simeq 0.5$~fm.

In all cases, we have collected 300 well 
decorrelated configurations, on which topological quantities have been 
measured both by staggered spectral projectors and, in order to make
a comparison, with a standard gluonic definition of the topological charge.
We note that statistics are not large, because the main purpose 
of our numerical simulations is to test the staggered definition
of spectral projectors
and not to make a precision study about topology in the pure gauge theory.

Concerning the gluonic definition, in this work we adopted the clover discretization:
\beq
\begin{gathered}
Q_{\text{\textit{clov}}} = \frac{-1}{2^9 \pi^2} \sum_{x}\sum_{\mu\nu\rho\sigma=\pm 1}^{\pm4} \epsilon_{\mu\nu\rho\sigma} \Tr\{\Pi_{\mu\nu}(x)\Pi_{\rho\sigma}(x)\}.
\end{gathered}
\eeq
As for the smoothing method, we decided to apply the standard cooling procedure, performing 80 cooling sweeps for each configuration 
(the topological susceptibility was stable already after 30 sweeps).
The cooled topological charge has been further rounded to the closest
integer following the procedure described in Refs.~\cite{DelDebbio:2002xa,Bonati:2015sqt} and then used to compute the gluonic definition of
topological susceptibility via
\beq
\chi_{\text{\textit{gluo}}} = \frac{\braket{Q^2_{\text{\textit{gluo}}}}}{V}.
\eeq

\subsection{Spectral determination of $\chi$ at $T=0$}\label{subsec_cont_limit_chi}

\begin{figure}[!htb]
\includegraphics[scale=0.5]{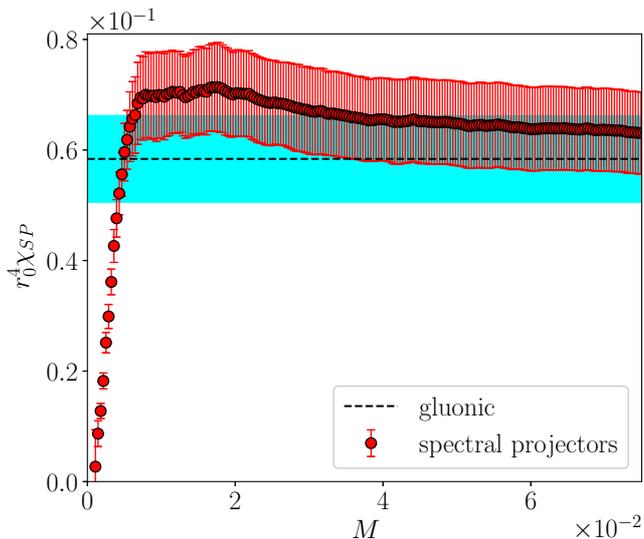}
\caption{Behavior of $\chi_{\text{\textit{SP}}}$ as a function
of the bare cut-off mass $M$, compared to the value of $\chi_{\text{\textit{gluo}}}$, for $\beta = 6.25$. Both susceptibilities are expressed in units of $r_0^{-4}$ while $M$ is reported in lattice units.}
\label{chi_vs_M}
\end{figure}

To start with,
in Figure~\ref{chi_vs_M} we show the 
topological susceptibility $\chi_{SP}$ obtained via spectral projectors for 
$\beta = 6.25$ and different values of the bare
cut-off mass $M$, comparing it with the gluonic determination on the 
same sample of configurations.
We observe an approximate plateau  
in a wide range of $M$, where $\chi_{SP}$ is in good agreement with the 
gluonic definition. Such a plateau is not required a priori, however
it is reasonable to expect it: the cut-off mass $M$ filters
away fluctuations at the UV scale, in particular $M^{-1}$ can be 
viewed as the analogous of the smoothing radius for 
smoothing techniques, so that the appearance of the plateau
is the signal of a well defined separation between the UV scale and the 
physical scale of topological excitations.

\begin{figure}
\includegraphics[scale=0.5]{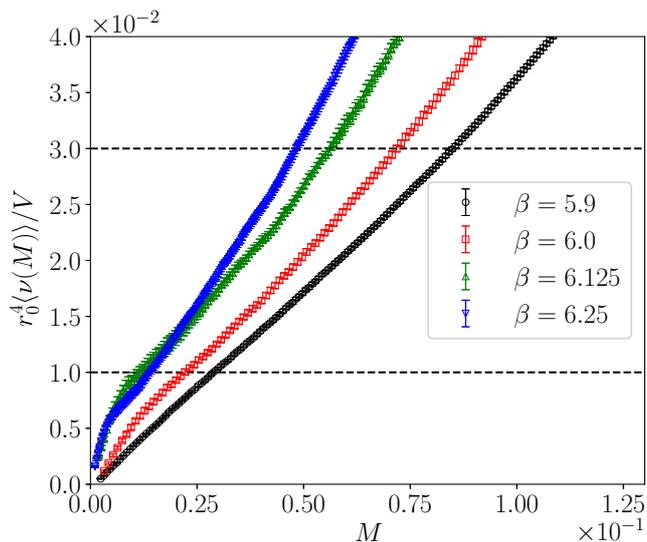}
\caption{Behavior of $\braket{\nu(M)}/V$ as a function of
the bare cut-off mass $M$ for 4 different values of the lattice spacing. The mean number of modes is expressed in units of $r_0^{-4}$ while $M$ is in lattice units.}
\label{nu_vs_M}
\end{figure}

In order to extrapolate $\chi_{\text{\textit{SP}}}$ towards the continuum, 
we considered determinations at fixed values of the 
renormalized mass $M_R$.
To keep $M_R$ fixed, 
we measured the dependence of $\braket{\nu(M)}/V$ on $M$ so that, 
once fixed a particular value of the mode density, we could find, 
for each $\beta$, the value of $M$ corresponding
to the same renormalized mass $M_R$. Fig.~\ref{nu_vs_M} 
illustrates this procedure for two different values of $M_R$ 
employed for the continuum extrapolation. 
In Table~\ref{tab_chi_vs_a} we report, for each $\beta$, the corresponding 
value of the lattice spacing, the volume in lattice units and the measures 
of the topological susceptibility obtained with spectral projectors and with 
the gluonic definition.

\begin{figure}[!htb]
\includegraphics[scale=0.5]{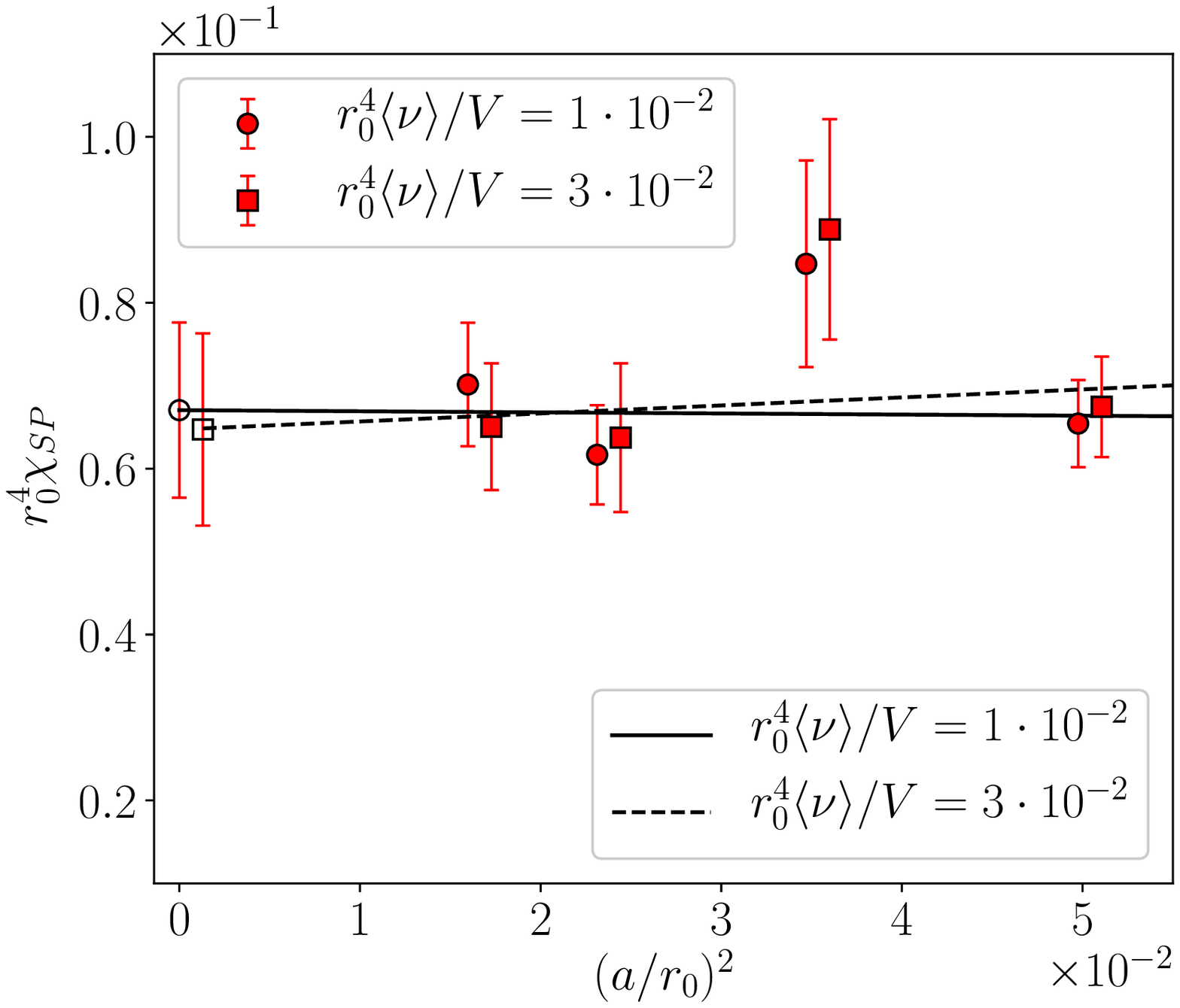}
\includegraphics[scale=0.5]{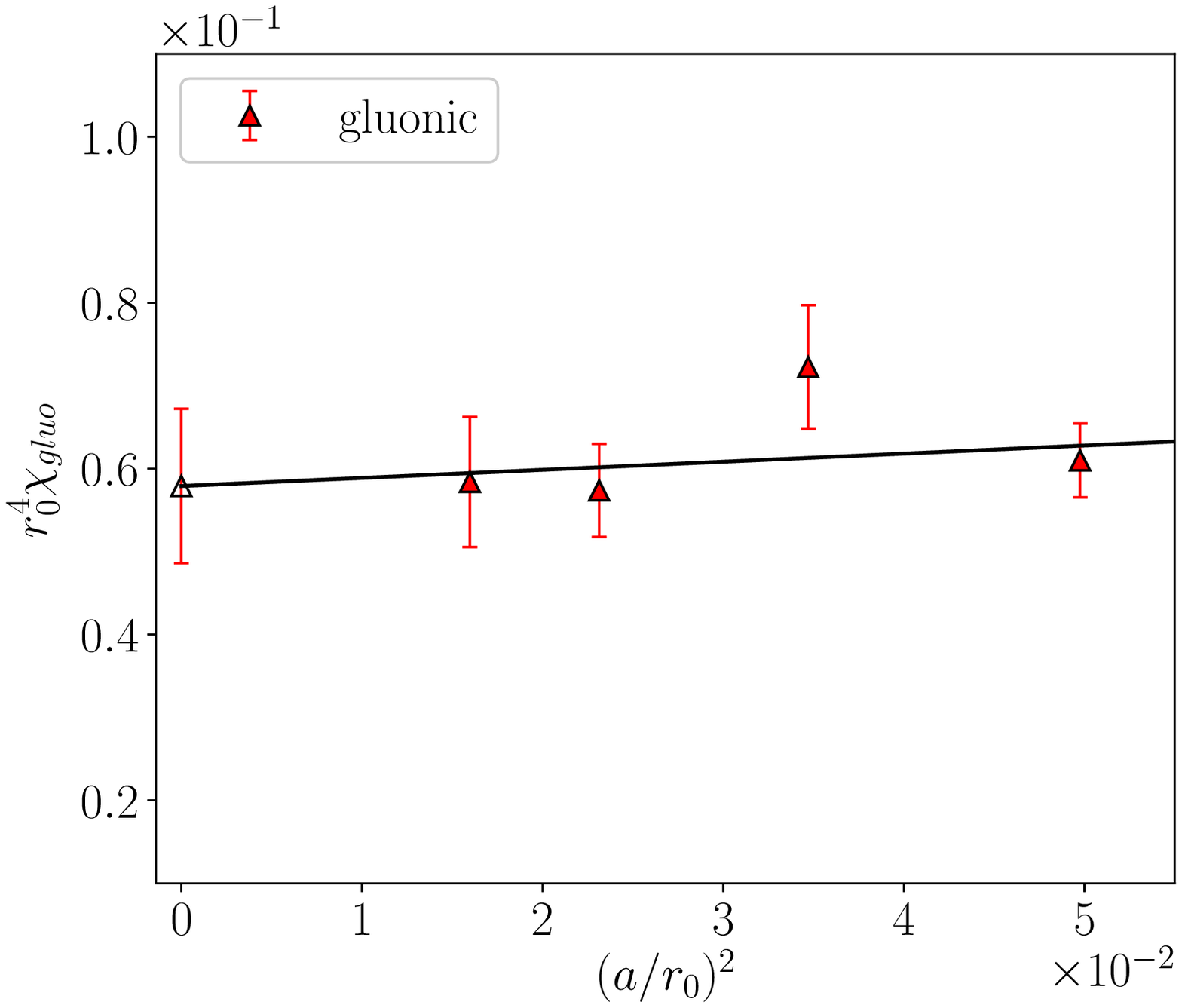}
\caption{Extrapolation towards the continuum of $\chi_{\text{\textit{SP}}}$ and $\chi_{\text{\textit{gluo}}}$ at $T=0$.}
\label{cont_limit_chi_plot}
\end{figure}

\begin{table}
\begin{center}
\begin{tabular}{ | c | c || c | c | c | c |}
\hline
& & & & & \\[-1em]
$\beta$ & $V$ & $(a/r_0)^2$ & $r_0^4\chi_{\text{\textit{SP}}}(M_1)$ & $r_0^4\chi_{\text{\textit{SP}}}(M_2)$ & $r_0^4\chi_{\text{\textit{gluo}}}$ \\
\hline
& & & & & \\[-1em]
5.9   & $16^4$ & 0.0498 & 0.0654(53) & 0.0675(61) & 0.0610(44) \\
6.0   & $16^4$ & 0.0347 & 0.085(12)  & 0.089(13)  & 0.0722(75) \\
6.125 & $16^4$ & 0.0231 & 0.0617(60) & 0.0637(90) & 0.0574(56) \\
6.25  & $24^4$ & 0.0160 & 0.0701(74) & 0.0651(76) & 0.0584(78) \\
\hline
\end{tabular}
\end{center}
\caption{Determinations of $\chi_{\text{\textit{SP}}}$ and $\chi_{\text{\textit{gluo}}}$ for each $\beta$. The lattice spacings in units of the Sommer parameter $r_0$ were taken from Ref.~\cite{guagnelli_lat_spacing_r_0}. The values of the renormalized cut-off masses $M_1$ and $M_2$ correspond, respectively, to $r_0^4 \braket{\nu}/V = 1 \cdot 10^{-3}$ and $3 \cdot 10^{-3}$. Assuming Eq.~(\ref{eq_for_cut_off_M}) and the values of $r_0$ and $\Sigma_R$ measured respectively in \cite{Sommer:2014mea} and \cite{Giusti:1998wy}, their values in physical units are $M_1\simeq 33$ MeV and $M_2\simeq 98$ MeV.}
\label{tab_chi_vs_a}
\end{table}
As shown in Table \ref{tab_chi_comparison}, the continuum value of $\chi$ obtained with spectral projectors is independent of the choice of $M_R$ and well compatible with the gluonic determination within the errors. For the sake of completeness, we also report determinations of $\chi$ obtained by other fermionic
methods, in particular using the overlap 
operator and Wilson spectral projectors. They all agree, within errors, 
with our staggered spectral determination. Fig.~\ref{cont_limit_chi_plot} 
shows the extrapolation towards the continuum both for $\chi_{\text{\textit{SP}}}$ and $\chi_{\text{\textit{gluo}}}$. Lattice artifacts have a slight dependence on the cut-off mass $M_R$, which is however well contained 
within errors and comparable in magnitude to that 
affecting the gluonic definition. 
This is similar to what happens in the case of  
Wilson spectral projectors~\cite{luscher_3,Cichy_wilson_spectral_quenched}.
\begin{table}[!htb]
\begin{center}
\begin{tabular}{ | c || c |}
\hline
& \\[-1em]
Method & $r_0^4 \chi$ \\
\hline
& \\[-1em]
Stag. Spectral Proj. $M_1$               & 0.067(11)  \\
Stag. Spectral Proj. $M_2$               & 0.065(12)  \\
Gluonic Def. $+$ Cooling                 & 0.058(9) \\
Overlap Operator~\cite{DelDebbio:2004ns} & 0.059(3)   \\
Wilson Spectral Proj.~\cite{luscher_3}   & 0.067(3)   \\
\hline
\end{tabular}
\end{center}
\caption{Comparison between various determinations of $\chi$ 
in the continuum for the pure gauge $SU(3)$ theory. The values of $M_1$ and $M_2$ are the same reported in  Table~\ref{tab_chi_vs_a}.} 
\label{tab_chi_comparison}
\end{table}
\subsection{Spectral determination of $b_2$ at high $T$}

The next-to-leading coefficient in the $\theta$-expansion of the free energy energy is (see Eqs.~(\ref{theta_dep_vacuum_energy}) and (\ref{cont_expression_b_2n}))
\beq
b_2 = -\frac{1}{12}\frac{\braket{Q^4}-3\braket{Q^2}^2}{\braket{Q^2}}.
\eeq
The gluonic discretization simply yields:
\beq
b_2^{\text{\textit{gluo}}} = -\frac{1}{12}\frac{\braket{Q_{\text{\textit{gluo}}}^4}-3\braket{Q_{\text{\textit{gluo}}}^2}^2}{\braket{Q_{\text{\textit{gluo}}}^2}}.
\eeq
Instead, from Eq.~(\ref{b_2n_stag_spectr_proj}) we get the staggered spectral expression:
\beq
\label{b2_sppr}
b_2^{\text{\textit{SP}}} &=& -\frac{1}{2^d}\frac{1}{12} \frac{\braket{\Tr\{\mathbb{P}_M\}}}{\braket{\Tr{\{\Gamma_5 \mathbb{P}_M \Gamma_5 \mathbb{P}_M \}}}} \cdot \\
&& \cdot \frac{\braket{\Tr{\{\Gamma_5 \mathbb{P}_M\}^{4}}} -3\braket{\Tr{\{\Gamma_5 \mathbb{P}_M\}^{2}}}^2}{\braket{ \Tr{\{\Gamma_5 \mathbb{P}_M\}^2 } }} \ .
\nonumber
\eeq
The measure of $b_2$ at zero temperature 
requires in general quite large statistics, because
it is necessary to detect deviations from gaussianity
of the topological charge distribution $P(Q)$, which 
are small~\cite{DelDebbio:2002xa, DElia:2003zne, Giusti:2007tu, Ce:2015qha, Bonati:2015sqt, Bonati:2016tvi} 
and become less and less visible as the lattice volume is increased. 
For this reason, we decided to test the numerical determination
of $b_2$ via spectral projectors in the high-temperature, deconfined
phase of the $SU(3)$ pure gauge theory, since in that regime 
its value is larger than in the $T = 0$  case,
approaching the prediction $b_2 = -1/12$ from the
Dilute Instanton Gas Approximation (DIGA),
while at the same time the width of the distribution (proportional
to the topological susceptibility) is 
smaller~\cite{pisa_work_b2_high_T}. 
We have considered, in particular, a determination at 
$\beta = 6.305$ on a $30^3 \times 10$ lattice, corresponding
to a temperature $T \simeq 338$~MeV~$\simeq 1.145~T_c$, 
for which a determination
of $b_2$ by the gluonic method has been already reported in 
Ref.~\cite{pisa_work_b2_high_T}. 

\begin{figure}
\includegraphics[scale=0.5]{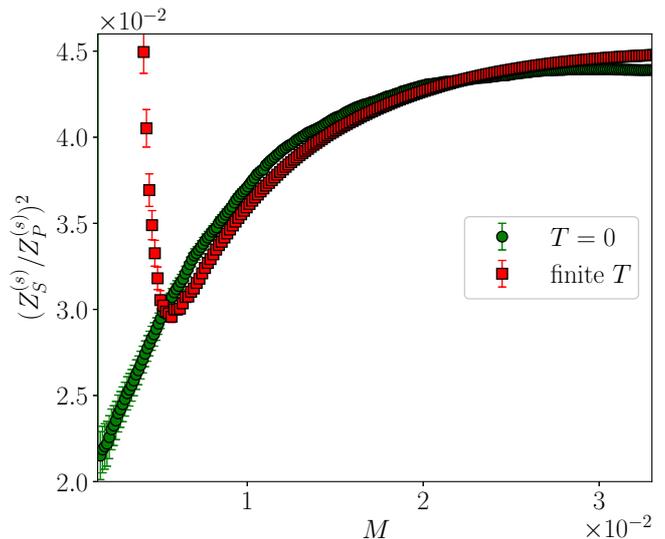}
\caption{Behavior of $(Z_S^{(\text{\textit{s}})}/Z_P^{(\text{\textit{s}})})^2$ as a function of $a M$ for two 
simulations performed at the same value of the bare coupling
($\beta = 6.305$, corresponding to $a=0.12\, r_0$) 
but on two different lattice sizes, 
$V=30^3 \times 10$ and $V=30^4$, corresponding respectively
to $T\sim338$ MeV and to an approximation of $T = 0$
(actually, $T \sim 130$~MeV, which is however deep in the confined phase).
}
\label{Z_S_Z_P_temperature_comparison}
\end{figure}

For a finite temperature implementation of the spectral projectors
method, an ambiguity could emerge as to whether the first ratio 
in Eq.~(\ref{b2_sppr}), corresponding to the multiplicative 
renormalization $(Z_P^{(\text{\textit{s}})}/Z_S^{(\text{\textit{s}})})^2$, should be computed in the finite 
$T$ simulation or instead at zero $T$. In principle, renormalization
constants should be independent of infrared (IR) conditions such as the
temperature scale. In order to check that, we have computed
the ratio both from the finite temperature simulation and from 
a dedicated simulation on a symmetric $30^4$ lattice 
at the same value of $\beta$: results are shown and compared in 
Fig.~\ref{Z_S_Z_P_temperature_comparison}, where it clearly
appears that, apart from the lowest values of 
$M$, for which the sensibility to IR conditions is large,
the two determinations are in reasonable agreement with each other.

\begin{figure}[!htb]
\includegraphics[scale=0.5]{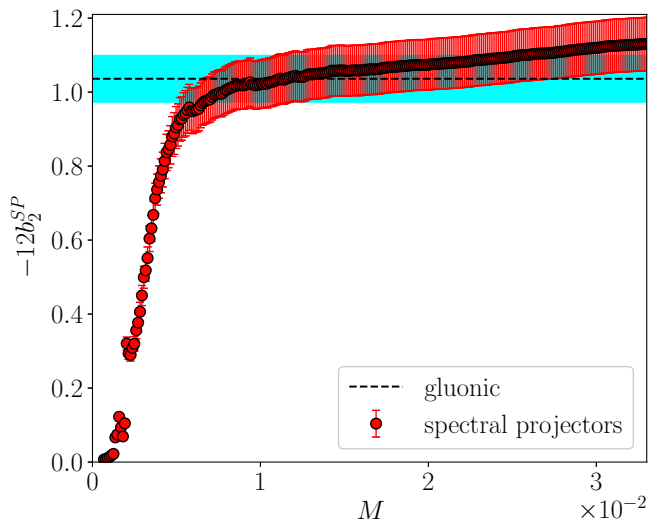}
\caption{Behavior of $- 12\, b_2^{\text{\textit{SP}}}$ at $T\sim338$ MeV as a function
of the bare cut-off mass $a M$, compared to the gluonic definition 
$b_2^{\text{\textit{gluo}}}$ determined on the same configuration
sample.
}
\label{b_2_vs_M_fig_beta_6.305}
\end{figure}

Finally, in Fig.~\ref{b_2_vs_M_fig_beta_6.305}, 
we show results obtained for $b_2^{\text{\textit{SP}}}$, and 
using the zero temperature renormalization constants, 
as a function of the bare cut-off mass $M$. In this case we do not 
fix a particular value of $M$, since we have data at a single 
value of $\beta$, hence 
we do not aim at performing the continuum extrapolation; however we notice
that results are in good agreement, over a wide range $M$, with the
gluonic determination of $b_2$ performed on the same configuration
sample, as well as with the determination of 
Ref.~\cite{pisa_work_b2_high_T} with the same
$\beta$ and lattice size ($ - 12\, b_2 = 1.10(7)$).

\section{Discussion and Conclusions}
\label{conclusions}
In this work we have defined the extension of the spectral projectors
method to the case of staggered fermions. Despite the different
patterns of chiral symmetry breaking, the final formula for the 
topological susceptibility, Eq.~(\ref{stag_sp_topo_susc}),
turns out to be practically identical to the one for Wilson fermions,
when the proper staggered discretization of the $\gamma_5$ operator 
and the fourfold degeneracy of staggered fermions are taken into account.
Moreover, the method has been extended to all higher-order cumulants
of the topological charge distribution, which enter the Taylor 
expansion in $\theta$ of the free energy density.
The method has then been tested in the pure $SU(3)$ gauge theory,
both at zero temperature and, for the fourth order cumulant, at finite 
$T$, with results in agreement with previous results in the literature
obtained by other fermionic or gluonic definitions of the
topological charge. 

Corrections to the continuum
limit turn out to be of the same order of magnitude as those
observed for the gluonic definition. Notice that the situation
could be quite different for full QCD simulations with light
quark masses: in this case
lattice artifacts for topological observables 
are in general significantly larger
than in the quenched theory, and the results 
of Ref.~\cite{athenodorou_wilson_spectral_fullQCD} show that 
the use of spectral projectors 
can lead to 
a strong reduction of these corrections.
This is actually the main reason of our interest 
in spectral projectors, in view of future applications to the 
study of topological properties of high-$T$ QCD with staggered fermions,
where the impact of corrections to the continuum limit 
is particularly significant~\cite{Bonati:2015vqz,Petreczky:2016vrs,Borsanyi:2016ksw,Burger:2018fvb,Bonati:2018blm}.

The origin of this different behavior may lie in the fact that 
in full QCD lattice artifacts for topological quantities are mostly driven
by the bad chiral properties of the fermion determinant used for Monte Carlo
sampling, which is not capable of properly suppressing topological
excitations, since they are not associated with exact zero modes
as in the continuum. For this reason, the adoption of a fermionic
definition of topology based on the same discretization
used for Monte Carlo sampling could in principle improve the situation,
since the same operator failing to suppress topological 
excitations is used to detect them. 
\acknowledgments
We thank A.~Athenodorou, F.~D'Angelo and A.~Todaro for useful discussions. Numerical simulations have been performed on the MARCONI machine at CINECA, based on the agreement between INFN and CINECA (under projects INF18\_npqcd and INF19\_npqcd).
%
%

%
\end{document}